\documentclass[showpacs,preprint,preprintnumbers,amsmath,amssymb,superscriptaddress,floatfix]{revtex4}

\usepackage{graphicx}
\usepackage{latexsym}
\usepackage{epsfig}
\usepackage{epstopdf}
\usepackage{textcomp}

\begin{document}

\title{Two-color modulation transfer spectroscopy}

\author{A. P\'{e}rez Galv\'{a}n$^*$, D. Sheng, Luis A. Orozco}
\affiliation{Joint Quantum Institute, Department of Physics,
University of Maryland and National Institute of Standards and Technology, College Park, MD 20742-4100, USA.}
\author{Y. Zhao}
\affiliation{State Key Laboratory of Quantum Optics and Quantum
Optics Devices, College of Physics and Electronics Engineering,
Shanxi University, Taiyuan 030006, China}
\date{\today}
 \email{apg@umd.edu}

\begin{abstract}
{We present two-color modulation transfer spectroscopy as a tool for precision
studies of atomic properties of  excited states. The bi-colored
technique addresses a narrow set of velocity groups of a thermal
atomic vapour using a two-step transition to ``burn a hole" in the
velocity distribution. The resulting spectrum presents sub-Doppler
linewidths, good signal to noise ratio and the trademark sidebands
that work as an \textit{in situ} ruler for the energy spacing
between atomic resonances. The spectra obtained can be used for
different applications such as measurements of energy splittings or
stabilization of laser frequencies to excited atomic transitions.}
\end{abstract}


\maketitle

\section{Introduction}

Spectroscopic studies of hyperfine manifolds in alkalies, such as
measurements of energy separations, have benefitted by
the high precision of the experimental techniques available to
interrogate atoms \cite{demtroder96}. Their hydrogen-like
structure makes interpretation of experimental results straightforward in terms of electromagnetic fields
generated by the valence electron and nuclear moments. Precise
measurements in higher excited
states accessible through two-step transitions\cite{ perez08, lee07, chui05, marian04} have appeared in recent years. This has renewed 
interest in improving calculations in other
states where theoretical methods such as many-body perturbation
theory (MBPT) (see for example the recent book of W. R. Johnson \cite{johnsonbook} ) are yet to be tested against experimental results .

Precise measurements in excited states, beyond the first one, have
several experimental complications. Standard spectroscopic
techniques rely on the high population of atoms in the ground state
to guarantee a good signal to noise ratio of the fluorescence or
absorption of the atomic sample. In two-step transitions this is no
longer the case. The amount of population transferred to the
intermediate level, for reasonable powers of the lasers, tends to be small, and detectors at the desired frequency might no
be readily available.

We present in this paper two-color modulation transfer
spectroscopy as a tool for studies of atomic properties of higher excited states. The
method consist of two lasers (pump and probe) counter-propagating
through a thermal vapour. Before being directed to the interaction
region, one of the lasers is modulated.
The first step of the transition \textit{i.e.} the pump, connects
the ground state to a resonant intermediate state while the probe
scans over the desired energy manifold. We monitor the absorption of
the pump laser as a function of probe laser detuning. The non-linear
interaction of the lasers ``burns a hole" in the atomic ground state
population. The generated spectra presents sub-Doppler peaks
(sometimes called Lamb-Bennett dips) corresponding to the atomic
resonances with the trademark sidebands at their side. This
technique overcomes the two main inconveniences of direct absorption
of the probing laser \textit{i.e.} low signal to noise ratio and
non-availability of detectors at the desired wavelength.

We present two ladder systems in  $^{87}$Rb to illustrate the main
features of the technique and two different applications of the
modulation. We select the $5S_{1/2}\rightarrow5P_{1/2}\rightarrow6S_{1/2}$ and
the $5S_{1/2}\rightarrow5P_{3/2}\rightarrow5D_{5/2}$ ladder transitions to illustrate their different uses. The  amplitude of the probe laser is modulated for the first system
while the second system has its pump frequency modulated. The frequency modulation of the pump laser and good signal to noise ratio allows
us to lock the probe laser to the $5P_{3/2}\rightarrow5D_{5/2}$
excited atomic resonance. In this case the probe laser remains
modulation free. This is highly desired since the electronic
modulation of the laser itself can carry unwanted effects such as
sidebands at higher or lower frequencies as well as bandwidth
problems. The method we are presenting is, of course, not limited to
these two cases and can be extended to other atomic levels.

The organization of the paper is as follows: section II contains the
theoretical model, section III explains the experimental setup and
results, section IV has a summary of the precise measurements using this method, and section V presents the conclusions.
\section{Theoretical model}

We start with a three
level model that can show some of the qualitative features of the 
experimental spectra. We use a density matrix
formalism to describe a three level atom in ladder configuration
interacting with two lasers, one of which has sidebands. We model
our system as Doppler-free ignoring Zeeman sublevels to keep
it tractable. The experimental situation is more complex and  for quantitative analysis it is necessary to take into account those same effects that we are ignoring.

\begin{figure}
\leavevmode \centering
  \includegraphics[width=7.5cm]{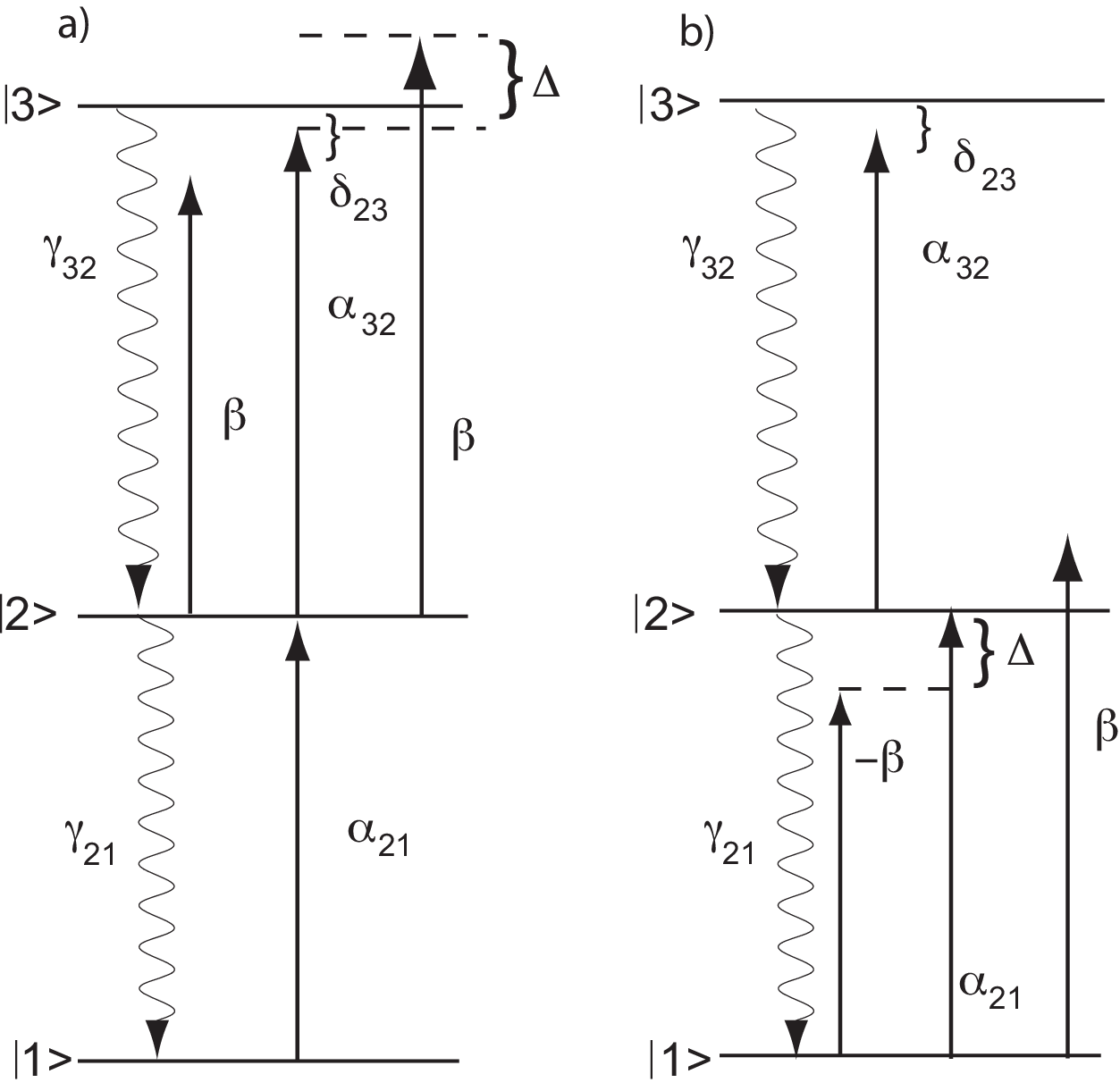}
  \caption{Theoretical model of two-step transition with (a) amplitude modulated probe and (b) frequency modulated pump.}
  \label{figure energy levels theory}
\end{figure}


Figure \ref{figure energy levels theory} shows our theoretical model.
We treat two cases. Fig \ref{figure energy levels theory} (a) is a
ladder type system with an amplitude modulated probe (AMP). Fig (b)
presents the same system except it has a frequency modulated pump
(FMP).

 The
intermediate and last levels are coupled by a single laser with three frequencies: a carrier
and two sidebands separated form the carrier by $\Delta$ (in MHz).
We represent the amplitude of the carrier by a Rabi frequency
$\alpha_{32}$ and the sidebands by a modulation depth $\beta$. The
ground and intermediate states are coupled by $\alpha_{21}$. The
detuning of the carrier between levels $|1\rangle$ and $|2\rangle$
is zero in the model as it is for our experiment and we let the detuning between levels
$|2\rangle$ and $|3\rangle$ vary as $\delta_{23}$. The total
population is normalized to unity. Fig. \ref{figure energy levels
theory} (b) follows the same nomenclature except that the sidebands
arise from frequency modulation and they appear in the pump laser
$\alpha_{21}$. For the FMP systems the sidebands have the
appropriate sign difference.

We have a set of nine linear equations for the slowly varying
elements of the density matrix $\sigma_{nm}$  after using the
rotating wave approximation with the sidebands rotating - one
clockwise, one counter clockwise - at a frequency $\Delta$ . The
equations are:

\begin{eqnarray*}
\lefteqn{\sum_{k}(\gamma_{kn}\sigma_{kk}-\gamma_{nk}\sigma_{nn})~+}&\\
  & & \frac{i}{2}\sum_{k}(\alpha_{nk}\sigma_{kn}-\sigma_{nk}\alpha_{kn})=\dot{\sigma}_{nm}~for~n=m,\nonumber\\
\lefteqn{[i(\Omega_{nm}-\omega_{nm})-\Gamma_{nm})]\sigma_{nm}~+}\\
  & & \frac{i}{2}\sum_{k}(\alpha_{nk}\sigma_{km}-\sigma_{nk}\alpha_{km})=\dot{\sigma}_{nm}~for~n\neq m,\nonumber
\end{eqnarray*}
where $\omega_{nm}=(E_{n}-E_{m})/\hbar$ is the transition frequency, and
$\Omega_{nm}=-\Omega_{mn}$ is the laser frequency connecting the
levels. The damping rate is given by:
\begin{eqnarray*}
\Gamma_{nm}=\frac{1}{2}\sum_k(\gamma_{nk}+\gamma_{mk}),
\end{eqnarray*}
and $\alpha_{21}=\alpha^{0}_{21}(1+\beta e^{i\Delta t}-\beta
e^{-i\Delta t})$ for the FMP system and
$\alpha_{32}=\alpha^{0}_{32}(1+\beta e^{i\Delta t}+\beta e^{-i\Delta
t})$ for the AMP system. The time dependence of the Rabi frequency
makes the standard approach of obtaining the steady state solution
of the system not feasible. Instead, we use a Floquet basis
expansion of the density matrix \cite{wong04} to solve our system of
equations. We replace each of the slowly rotating elements of the
density matrix by:

\begin{eqnarray*}
\sigma_{nm}(t)=\sum^{p}_{k=-p}\sigma^{(k)}_{nm}e^{ik\Delta t},
\end{eqnarray*}
where $\sigma^{(k)}_{nm}$ is the Fourier amplitude of the component
oscillating at $k\Delta$. The system is now a series of $2p+1$
coupled equations for some large $p$ that have to be solved
recursively. It is necessary to set $\sigma^{(p)}_{nm}=0$ for some
$p$ to cut off the infinite number of coupled equations. By solving
the equations in terms of their predecessors
we can extract $\sigma_{12}(t)$. 

\begin{figure}
\leavevmode \centering
  \includegraphics[width=7.5cm]{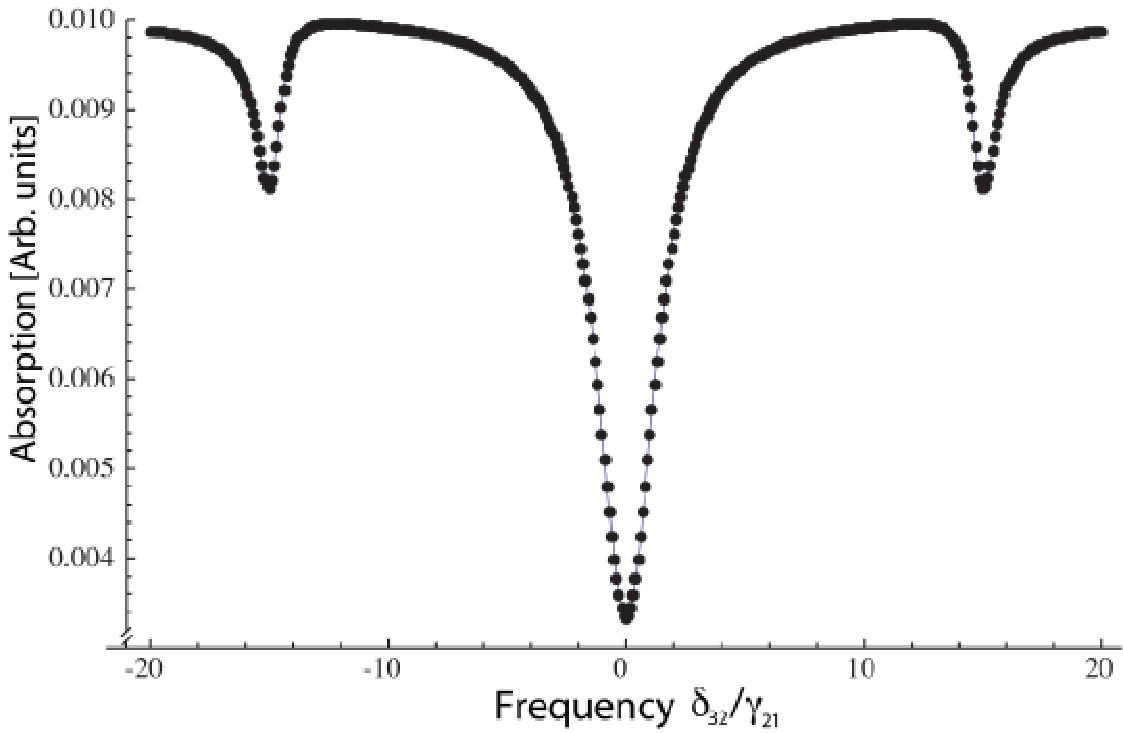}
  \caption{Numerical solution of the absorption of the pump laser as a function of the normalized detuning
 of the probe laser in units of $\gamma_{21}$. The parameters are (in units of $\gamma_{21}$) :
 $\beta=1/3$, $\alpha_{21}=1/100$, $\alpha^{0}_{32}=1/4$, $\gamma_{32}=1/2$, and $\Delta=15$}
  \label{figure absorption AM}
\end{figure}

\begin{figure}
\leavevmode \centering
  \includegraphics[width=7.5cm]{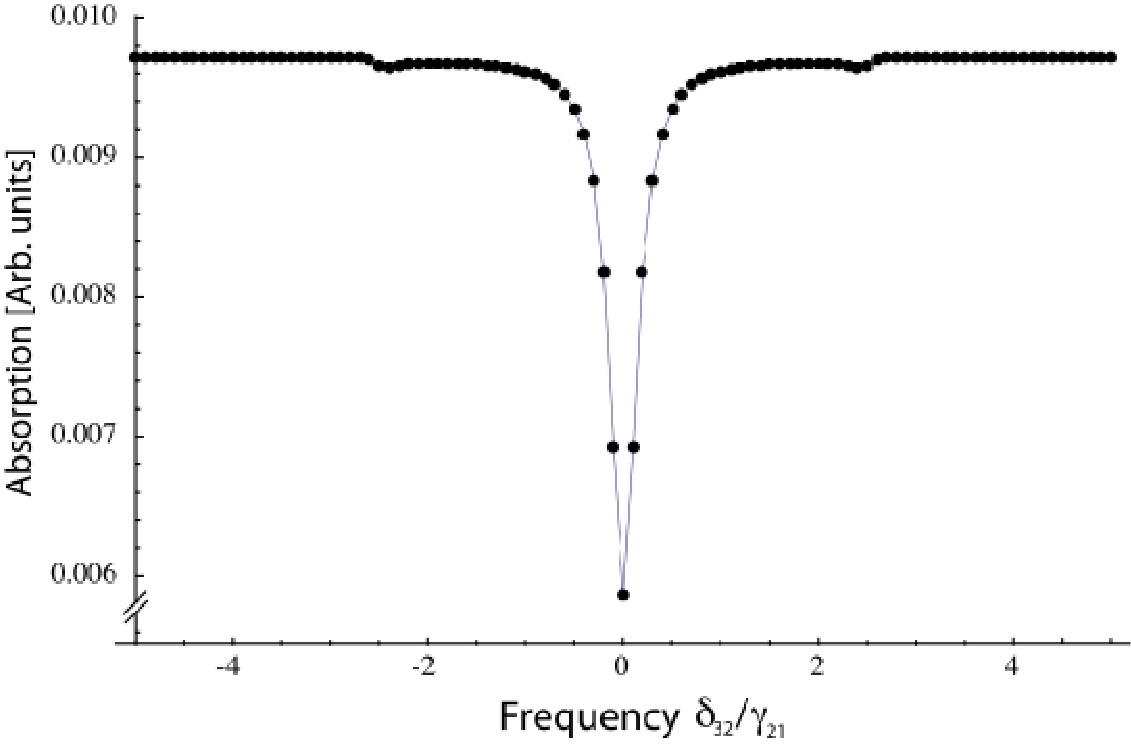}
  \caption{Numerical solution of the absorption of the pump laser as a function of the normalized
detuning of the probe laser in units of $\gamma_{21}$. The
parameters are (in units of $\gamma_{21}$) : $\beta=1/10$,
$\alpha^{0}_{12}=1/100$, $\alpha_{23}=1/4$, $\gamma_{32}=1/10$, and
$\Delta=2.5$}
  \label{figure absorption FM}
\end{figure}

For our experiment we are
interested in the terms  $\sigma^{(0)}_{12}$, $\sigma^{(-1)}_{12}$,
and $\sigma^{(1)}_{12}$ which are proportional to the absorption of
the first laser carrier and sidebands, respectively. We plot the
absolute value of the imaginary part as a function of $\delta_{23}$
to recover the absorption. This is necessary to take into account
the square-law nature of the photodiode. Our three level model
reproduces the resonance features of the absorption observed as the
second excitation goes into resonance for both AMP and FMP systems
(see Fig. \ref{figure absorption AM} and Fig. \ref{figure absorption
FM}, respectively). The demodulation of the FMP signal yields the
expected error-like feature shown in Fig. \ref{figure demodulated
absorption}.

\begin{figure}
\leavevmode \centering
  \includegraphics[width=7.5cm]{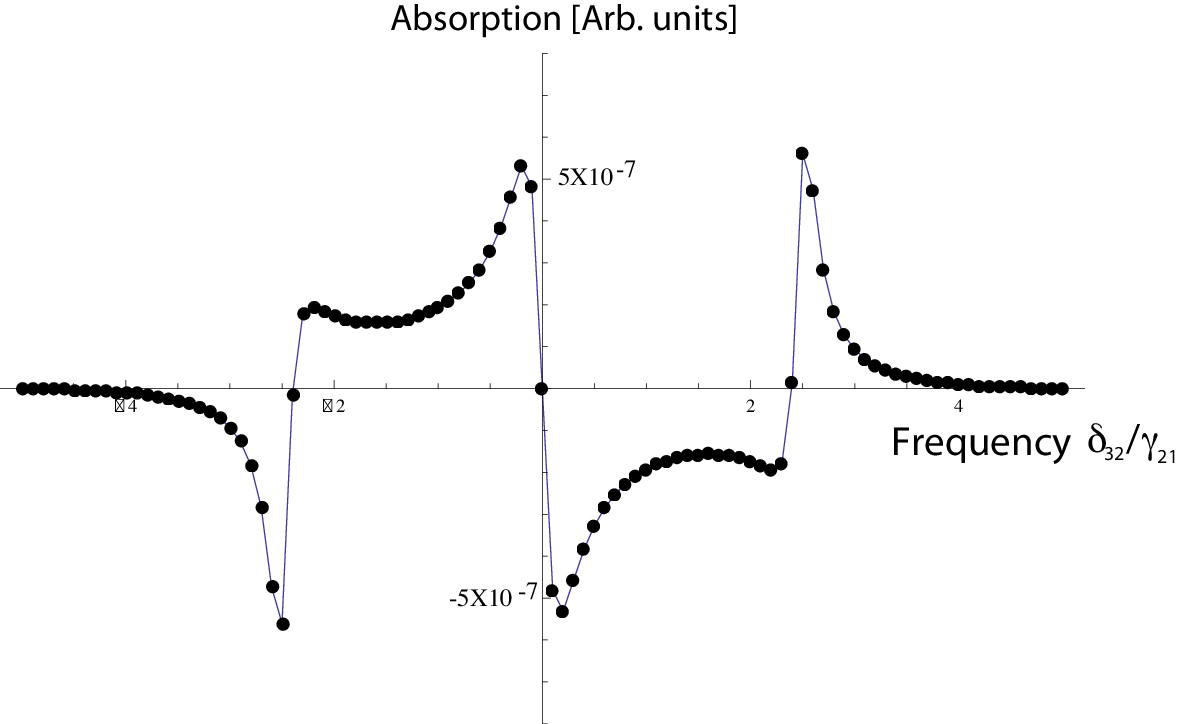}
  \caption{Numerical simulation of the demodulated absorption of the pump laser for the FMP system as a function of the normalized detuning of the probe
  laser in units of $\gamma_{21}$. The parameters are (in units of $\gamma_{21}$) : $\beta=1/10$,
 $\alpha^{0}_{12}=1/100$, $\alpha_{32}=1/4$, $\gamma=1/10$, and $\Delta=2.5$}
  \label{figure demodulated absorption}
\end{figure}

The size of the sidebands in our model depends on the modulation index (separation from resonance and strength), as well as the specific decay rates of the levels which set up the Rabi frequencies $\alpha_{ij}$ in the  AMP and FMP systems.

\section{Apparatus and method}

\begin{figure}
\leavevmode \centering
  \includegraphics[width=7.5cm]{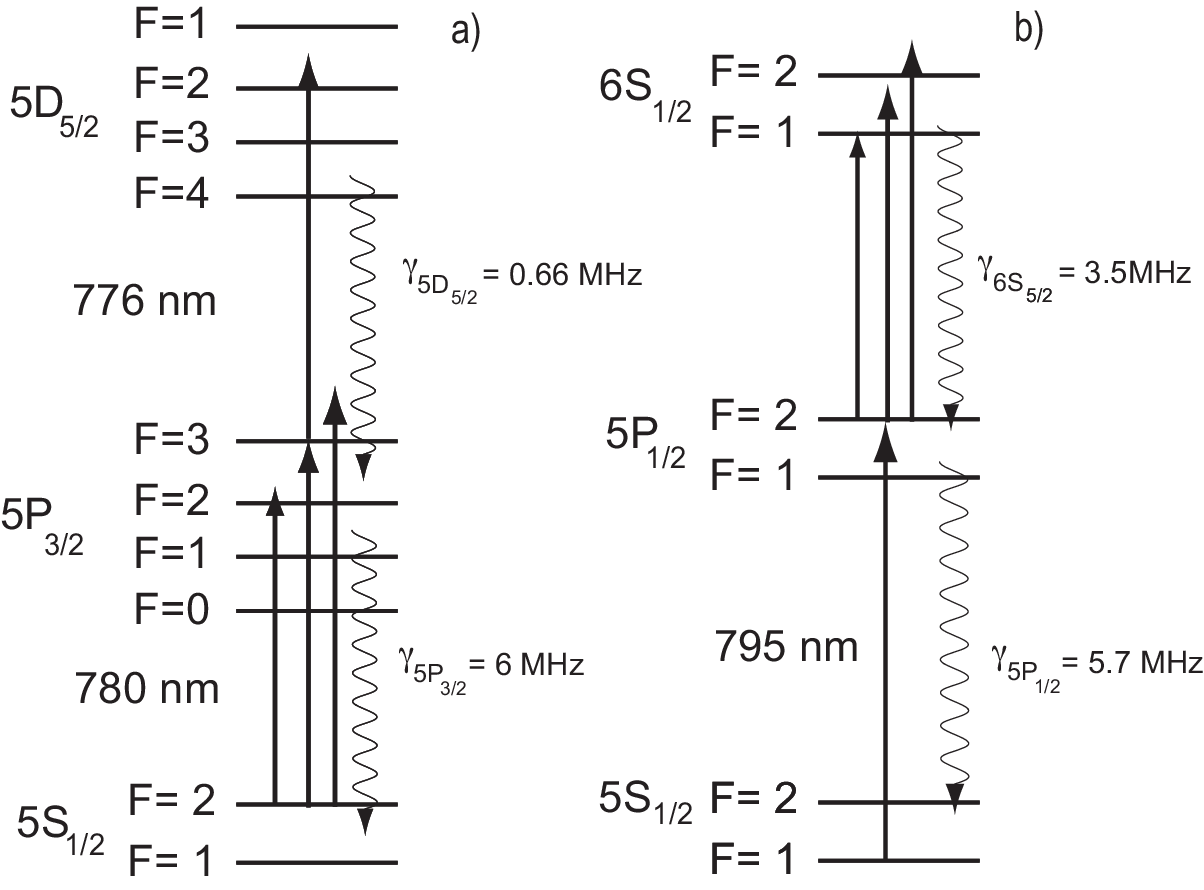}
  \caption{a) Relevant energy levels for the modulation of the pump atomic system.
 (b) Relevant energy levels for the modulation of the probe atomic system.}
  \label{figure energy levels}
\end{figure}

\begin{figure}
\leavevmode \centering
  \includegraphics[width=7.5cm]{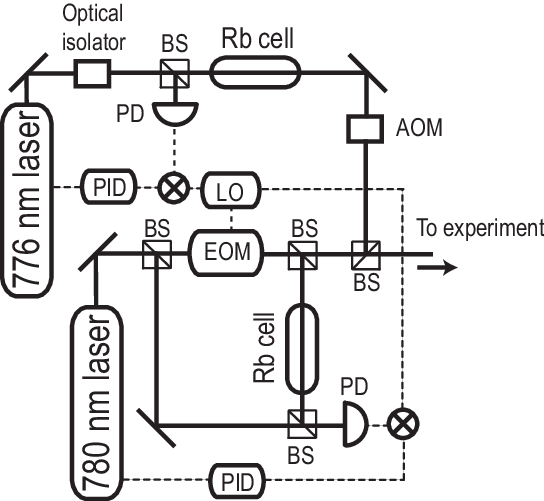}
  \caption{Block diagram of the experimental. Key for the figure PD: photodiode, AOM: acousto-optical modulator, LO: local oscillator, BS: beam splitter. }
  \label{figure experimental setup pump}
\end{figure}

Figure \ref{figure experimental setup pump} and Fig. \ref{figure
experimental setup probe} present block diagrams of the FMP and AMP
systems, respectively. A Coherent
899-01 Ti:sapphire laser with a linewidth of less than 100 kHz is the pump laser in both cases. A
small amount of laser power from the pump laser is frequency
modulated by a small bandwidth electro-optical modulator at
$\approx$15 MHz and sent to a glass cell filled with rubidium at
room temperature to lock the laser frequency to the $5P_{3/2}$
crossover line of the $F=1$ and $F=3$ hyperfine levels for the FMP
system and to the on resonance $F=1\rightarrow F=2$ transition of
the $5P_{1/2}$ level for the AMP system at 795 nm with a
Pound-Drever-Hall lock.

Level $|1\rangle$ in the AMP system corresponds to the lower
hyperfine state of the 5S$_{1/2}$ level ($F=1$) while $|2\rangle$ is
the highest hyperfine state of the 5P$_{1/2}$ level ($F=2$) of
$^{87}$Rb . The decay rate between the two levels is
$\gamma_{21}/2\pi=$ 5.7~MHz \cite{simsarian98}. We simplify the
hyperfine states of the $6S_{1/2}$ level to just one level with
decay rate $\gamma_{32}/2\pi=$ 3.5~MHz \cite{gomez05b}.

For the FMP system, 
the probe laser is an SDL
diode laser with a linewidth of 5 MHz at 776 nm. The lasers overlap
inside an independent rubidium glass cell at room temperature
wrapped in $\mu$-metal in lin-perp-lin polarization configuration.
Their $1/e^{2}$ power diameter of the laser beams is 1 mm. We scan the probe laser over
the $5D_{5/2}$ level hyperfine manifold and observe the absorption
of the pump laser as a function of the probe laser detuning using a
fast photodetector. We send the signal to a bias-T and record the DC
and demodulated AC components with a WaveSurfer digital oscilloscope
with an 8-bit resolution from Lecroy.

\begin{figure}
\leavevmode \centering
  \includegraphics[width=7.5cm]{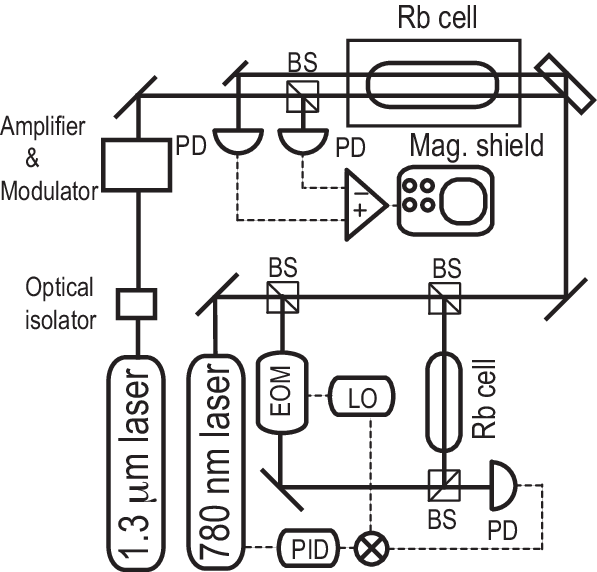}
  \caption{Block diagram of the experimental. Key for the figure PD: photodiode, LO: local oscillator, BS: beam splitter. }
  \label{figure experimental setup probe}
\end{figure}

We keep the power of the pump laser and the modulation depth fixed
to a value of 100 $\mu$W and $\beta=0.2$, respectively. We change
the power of the probe beam and observe its influence on the
spectra. It is possible to observe the resonant features of the
$5D_{5/2}$ hyperfine manifold with little as 100 $\mu$W of probe
power. Higher probe power increases the signal size and the width of
the features. Playing with the polarization and powers we also
observe EIT features \cite{banacloche95}. We restrict ourselves to a
space parameter where these very narrow features are absent.

\begin{figure}
\leavevmode \centering
  \includegraphics[width=7.5cm]{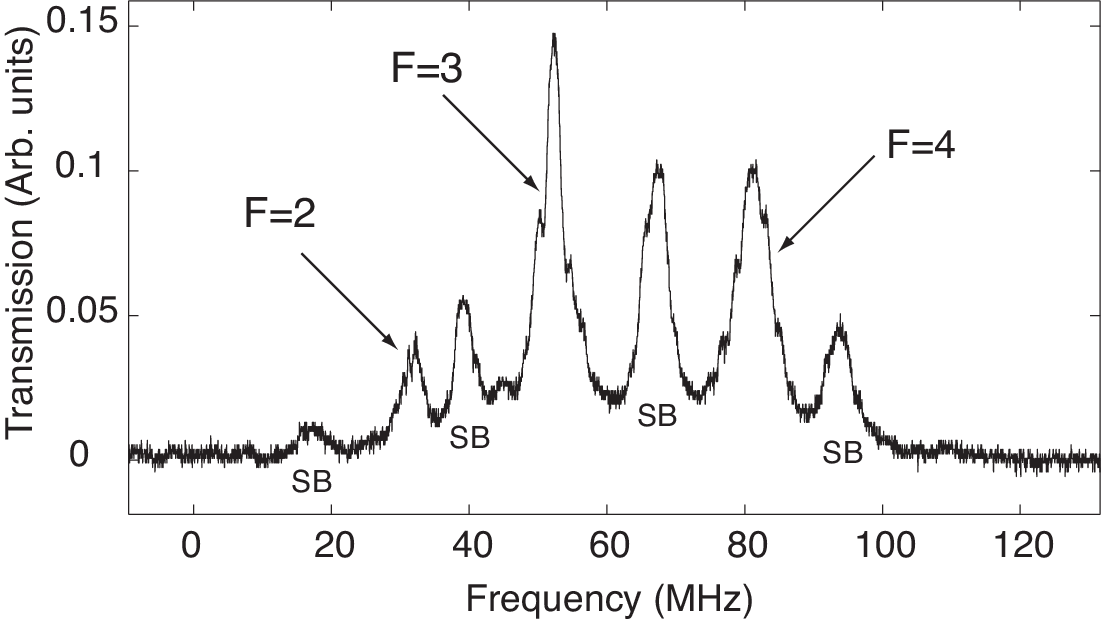}
  \caption{Experimental trace for FMP showing the DC component of the absorption of the 780 nm laser as a function of the probe laser detuning as it scans across the 5$D_{5/2}$ level in $^{87}$Rb. It presents the main resonances as well as the indicated sidebands (SB). The power of the probe
and pump beam are 4.3 mW and 100 $\mu$W, respectively. }
  \label{figure absorption with sidebands}
\end{figure}

\begin{figure}
\leavevmode \centering
  \includegraphics[width=7.5cm]{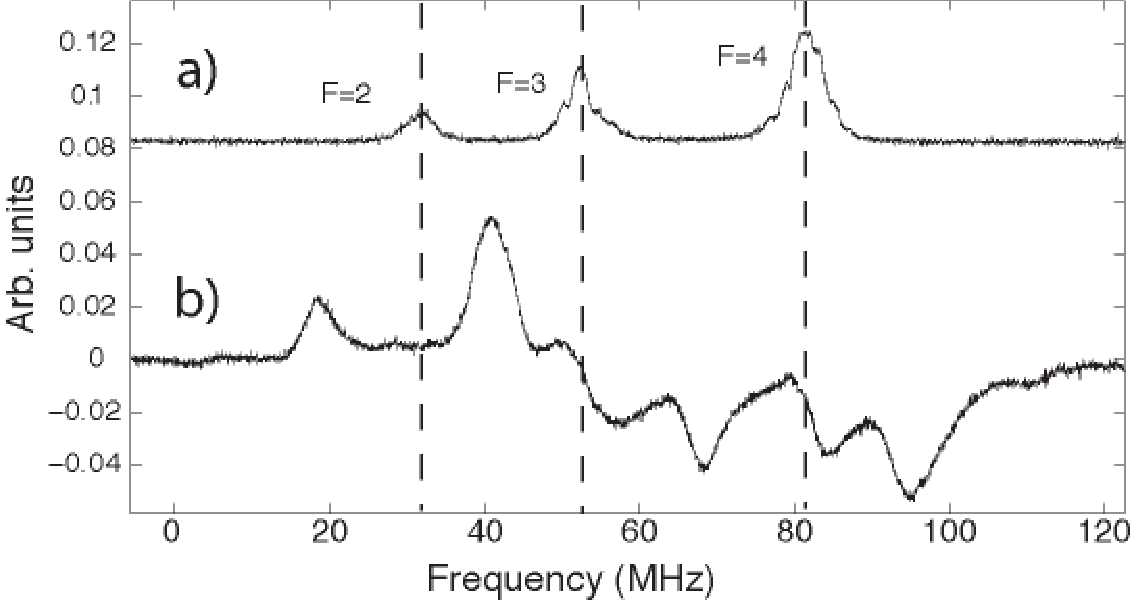}
  \caption{Experimental traces for FMP for the 5$D_{5/2}$ resonances in $^{87}$Rb of (a) absorption without sidebands and (b) demodulated absorption of 780 nm laser as a function of detuning of the 776
nm laser.  The power of the probe and pump beam are 4.3 mW and 100
$\mu$W, respectively.}
  \label{figure absorption}
\end{figure}

Figure \ref{figure absorption with sidebands} (FMP) and \ref{figure absorption} (FMP) show typical experimental traces of the absorption of
the 780 nm laser. The spectrum has been offset to zero transmission
for convenience. The first of these, Fig. \ref{figure absorption
with sidebands}, has the DC component of the absorption with the
sidebands appearing on both sides of the main resonances. No Doppler
background is observed for any of the experimental conditions
explored, showing that this is a Doppler free spectrum. Fig. \ref{figure absorption} (a) shows the lower hyperfine
states of the $5D_{5/2}$ level manifold with no sidebands for
clarity. Fig.~\ref{figure absorption}(b) has the demodulated AC component of the
absorption. The dashed lines identify the error-like features with
their corresponding hyperfine levels. We use this spectrum to
stabilize the frequency of the probe laser.

\begin{figure}
\leavevmode \centering
  \includegraphics[width=7.5cm]{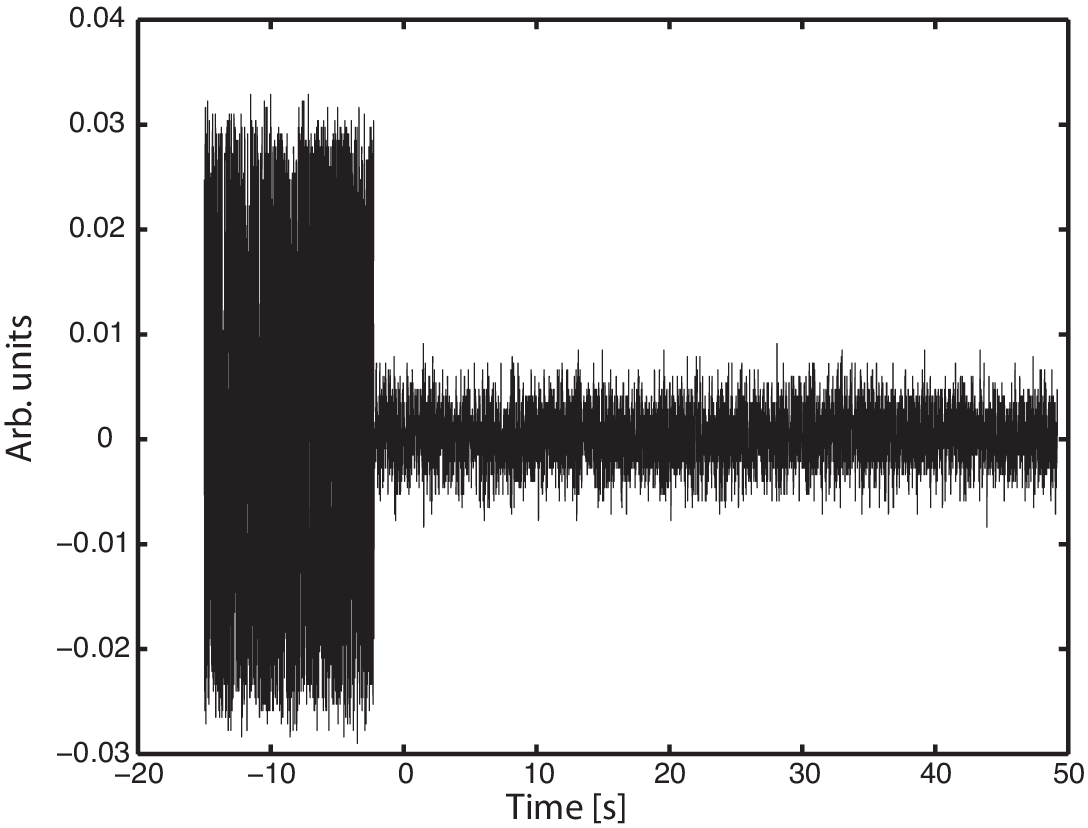}
  \caption{Fringe-side transmission of the SDL laser at 776 nm
through a confocal Fabry-Perot cavity. The reduction of the amplitude of the signal corresponds to the locking of the laser to the $5P_{3/2}\rightarrow5D_{5/2}$ excited atomic transition using FMP.}
  \label{figure error signal}
\end{figure}

We monitor the laser frequency of the probe beam
using a Coherent  confocal Fabry-Perot cavity with a free spectral
range of 1.5 GHz to test the performance of the laser lock. Fig. \ref{figure error signal} shows the
fringe-side transmission of the probe laser through the cavity. We
monitor the behavior of the laser before and after it has been
locked. The reduction of the frequency excursions is quite evident
as the laser is locked to the atomic resonance. Under normal
experimental conditions we have observed locking times of 30
minutes, and a significant reduction of the rms noise of more than a factor of seven.

A thick glass plate splits into two the main beam at 795 nm in  the AMP system before entering an independent rubidium vapor glass cell
inside a three layered magnetic shield. A grating
narrowed diode laser at 1.324 $\mu$m (from here on referred to as 1.3 $\mu$m laser) with a linewidth better than 500
kHz excites the second transition. We scan the frequency of the 1.3
$\mu$m laser over the hyperfine manifold of the $6S_{1/2}$ level. A fiber-coupled semiconductor
amplifier increases the power of the 1.3 $\mu$m laser before it goes
to a large bandwidth ($\approx$10 GHz) Electro-Optic Modulator (EOM)
that generates the sidebands.

The power of each 795 nm beam is approximately 10~$\mu$W with a
diameter of 1 mm. We operate in the low intensity regime to avoid
power broadening, differential AC stark shifts and line splitting
effects such as the Autler-Townes splitting. Both beams are
circularly polarized by a $\lambda/4$ waveplate. The counter
propagating 1.3 $\mu$m laser beam with a power of 4 mW and
approximately equal diameter overlaps one of the 795~nm beams. After
the glass cell an independent photodiode detects each 795~nm beam.
The outputs of the detectors go to a  differential amplifier to
reduce common noise. A digital oscilloscope records the output
signal for different values of modulation. Fig. \ref{figure whole
scan} shows an absorption spectrum of the 795~nm laser as a function
of the detuning of the 1.3 $\mu$m laser that shows the
signature sidebands of the technique. Fig. \ref{figure linear
regression} shows a plot of the distance between the sidebands as a
function of the modulation of the 1.3 $\mu$m laser. The sidebands
that appear on the absorption spectra provide \textit{in situ} calibration for the energy spacing of the hyperfine splittings. This effectively translates a measurement of energy spacings from the optical region to a much easier measurement in the radio frequency range.

\begin{figure}[t]
\leavevmode \centering
  \includegraphics[width=7.5cm]{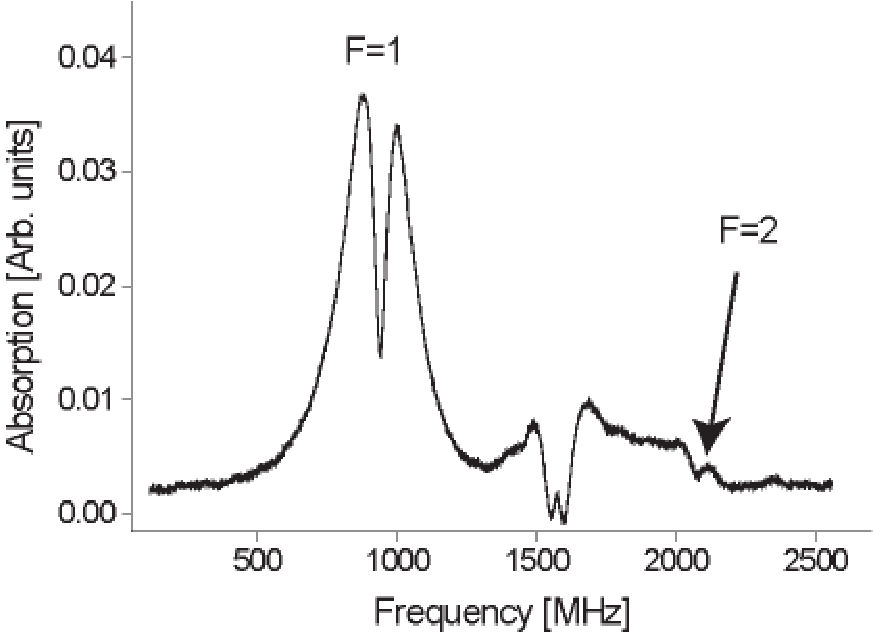}
  \caption{Absorption spectrum of the 795 nm pump laser as a function of the detuning of the modulated probe.  Non desired features appear in the spectrum when the experimental parameters are not under control.}
  \label{figure bad absorption}
\end{figure}

We observe a rich atomic behavior such as EIT and reversal of the peaks that depends on the power of the lasers, relative polarization and magnetic  field intensity (see for example Fig \ref{figure bad absorption}). This points towards a stringent control on experimental parameters for precision studies of energy separations.
\begin{figure}[t]
\leavevmode \centering
  \includegraphics[width=7.5cm]{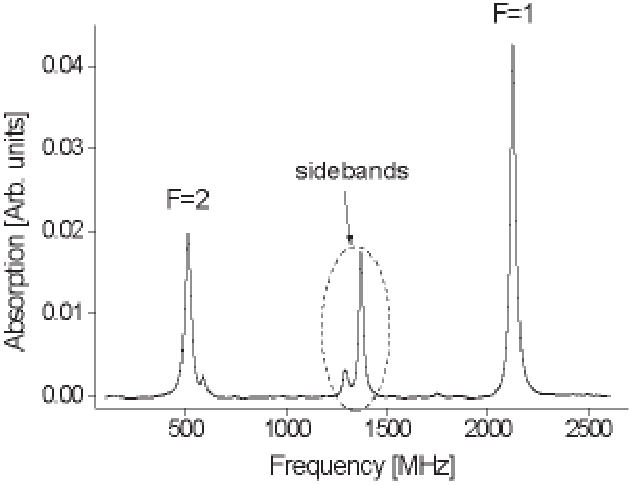}
  \caption{Absorption profile of the 6S$_{1/2}$, F=1 and F=2 hyperfine states of $^{87}$Rb with sidebands. The big sideband belongs to the F=1 peak.
 The small feature on the side of the F=2 peak corresponds to the second sideband of the F=1 peak. The glass cell is in a magnetic field of 0.37 G.}
  \label{figure whole scan}
\end{figure}

\begin{figure}[t]
\leavevmode \centering
  \includegraphics[width=7.5cm]{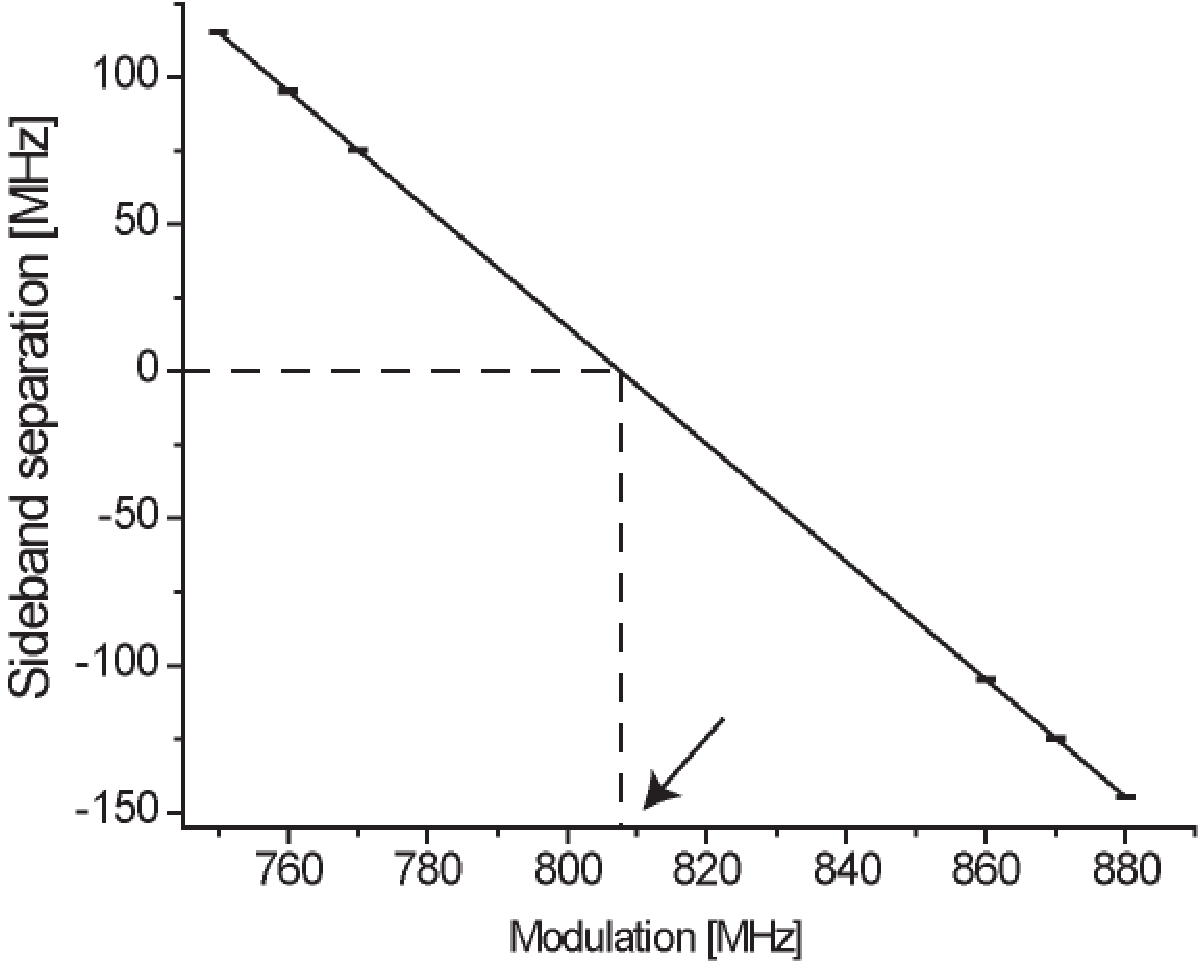}
  \caption{Plot of distance between sidebands as a function of modulation of the 1.3 $\mu$m laser for $^{87}$Rb. The arrow point to the value of the modulation that corresponds to the overlap of the sidebands and half the hyperfine splitting of the $6S_{1/2}$ level
hyperfine splitting.}
  \label{figure linear regression}
\end{figure}

\section{Precise measurements}

Table \ref{table theory and experiment} shows the values of the
magnetic dipole constants using relativistic MBPT \cite{safronova99}
with single double (SD) and single double partial triple (SDpT) wave
functions and values extracted from measurements of the hyperfine
splitting in other electronic states currently in the literature for
$J$=1/2 \cite{arimondo77,barwood91,marian04,marian05,chui05}. We
have not been able to find in the literature values for higher
levels with adequate precision to include them in the figure. The
agreement of the theory with the experiment, for $J$=1/2 levels, is
well within the 1\% level. The SDpT relativistic wave functions do
indeed improve the accuracy of the calculations of the single double
wave functions.

\begin{table}[h]
  \leavevmode \centering
   \begin{tabular}{l|ccc}
                 &  SD [MHz] & SDpT [MHz] & Experiment [MHz]\\\hline
   5$S_{1/2}$    &  642.6    & 1011.1     & 1011.910813(2) \cite{arimondo77}\\
   5$P_{1/2}$    &  69.8     & 120.4      & 120.499 (10) \cite{barwood91} \\
   6$S_{1/2}$    &  171.6    & 238.2      & 239.18(3) \cite{perez07,perez08}\\
   6$P_{1/2}$    &  24.55    & 39.02      & 39.11(3) \cite{marian04,marian05}\\
   7$S_{1/2}$    &  70.3     & 94.3       & 94.658(19) \cite{chui05} \\

   \end{tabular}
  \caption{Single Double (SD) and partial triple (SDpT) excitation calculated from \textit{ab intio} MBPT in Ref.  \cite{safronova99} and experiment magnetic dipole constants for the first $J$=1/2
   levels in $^{85}$Rb. (Adapted from Ref.~\cite{perez08}).}
  \label{table theory and experiment}
\end{table}

The accuracy of the 6$S_{1/2}$ measurement is high enough to extract a hyperfine anomaly \cite{perez07} in an excited state, which shows that the effect is independent of the n state of the level, as originally predicted by Bohr and Weiskopf \cite{bohr50}.

\section{Conclusions}

We have presented two-color modulation transfer
spectroscopy as reliable and simple method for studies of atomic
properties in excited states. The characteristic sidebands appearing
the spectra have the two-fold utility of working as an \textit{in
situ} ruler for measurements of energy separations or to lock the
frequency of a laser to an excited transition. The good quality of the data presented is due to
monitoring of the absorption of the pump beam
instead of direct absorption of the probe beam. The absorption of the pump beam (or lack thereof), is always
guaranteed since a vast amount of atoms are always in the ground
state and even small changes \textit{i.e.} excitation to the last step of the
transition, will be noticeable even for small powers of the pump
beam. In addition, the spectra does not show a Doppler background due to the lack of an
equilibrium thermal population in the intermediate state. It is the hope that the method will stimulate studies of
atomic properties of excited states and further push the
experimental precision and theoretical work in excited atomic
states.

\section{Acknowledgments} Work supported by NSF. A.P.G. would like to thank
E. Gomez for discussions on the subject of this article and P. Barberis for help on the theory of three level atoms.


\end{document}